\documentclass[sigconf]{acmart}

\makeatletter
\def\doclicense@image@file{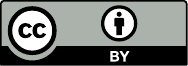}
\makeatother

\AtBeginDocument{%
  }

\copyrightyear{2026}
\acmYear{2026}
\setcopyright{cc}
\setcctype{by}
\acmConference[ICMR '26]{International Conference on Multimedia Retrieval}{June 16--19, 2026}{Amsterdam, Netherlands}
\acmBooktitle{International Conference on Multimedia Retrieval (ICMR '26), June 16--19, 2026, Amsterdam, Netherlands}
\acmDOI{10.1145/3805622.3810808}
\acmISBN{979-8-4007-2617-0/2026/06}

\usepackage{enumitem}
\usepackage{multirow}
\usepackage{booktabs}
\usepackage{makecell}
\usepackage{xcolor}
\usepackage{colortbl}

\begin{document}
\special{ps: /doclicense-CC-by-88x31.pdf}

\title{Mitigating Multimodal Inconsistency via Cognitive Dual-Pathway Reasoning for Intent Recognition}

\author{Yifan Wang}
\affiliation{%
  \institution{Hebei University of Science and Technology}
  \city{Shijiazhuang}
  \country{China}}
\email{2023114119@stu.hebust.edu.cn}

\author{Peiwu Wang}
\affiliation{%
  \institution{Hebei University of Science and Technology}
  \city{Shijiazhuang}
  \country{China}}
\email{peiwuwang@stu.hebust.edu.cn}

\author{Yunxian Chi}
\authornote{Yunxian Chi and Zhinan Gou are the corresponding authors.}
\affiliation{%
  \institution{Hebei University of Science and Technology}
  \city{Shijiazhuang}
  \country{China}}
\email{chiyunxian_hebtu@163.com}

\author{Zhinan Gou}
\authornotemark[1]
\affiliation{%
  \institution{Hebei University of Economics and Business}
  \city{Shijiazhuang}
  \country{China}}
\email{gouzhinan@hueb.edu.cn}

\author{Kai Gao}
\affiliation{%
  \institution{Hebei University of Science and Technology}
  \city{Shijiazhuang}
  \country{China}}
\email{gaokai@hebust.edu.cn}

\begin{abstract}
Multimodal Intent Recognition (MIR) aims to understand complex user intentions by leveraging text, video, and audio signals. However, existing approaches face two key challenges: (1) overlooking intricate cross-modal interactions for distinguishing consistent and inconsistent cues, and (2) ineffectively modeling multimodal conflicts, leading to semantic cancellation. To address these, we propose a novel Cognitive Dual-Pathway Reasoning (CDPR) framework, which constructs a stable semantic foundation via the intuition pathway and mitigates high-level semantic conflicts through the reasoning pathway, cooperatively establishing deep semantic relations. Specifically, we first employ a representation disentanglement strategy to extract modality-invariant and specific features. Subsequently, the intuition pathway aggregates cross-modal consensus using shared features for solid global representations. The reasoning pathway introduces an inconsistency perception mechanism, combining semantic prototype matching with statistical probability calibration to precisely quantify conflict severity, and dynamically adjusting the weights between both pathways. Furthermore, a multi-view loss function is adopted to alleviate modality laziness and learn structured features at different stages. Extensive experiments on two benchmarks show that CDPR achieves SOTA performance and superior robustness in mitigating multimodal inconsistency. The code is available at \url{https://github.com/Hebust-NLP/CDPR}.
\end{abstract}



\maketitle

% 78%
% Comparison between existing indiscriminate fusion paradigms and our proposed CDPR framework.
\begin{figure}[t!]
	\centering  
	\includegraphics[scale=.72]{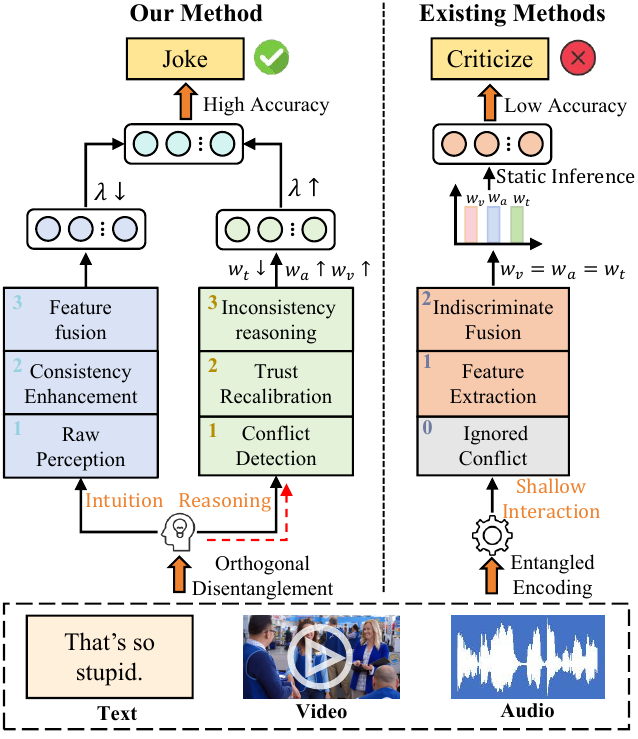}
	\caption{\label{example} Comparison of CDPR and Existing Paradigms.}
\end{figure}

\section{Introduction}
Multimodal Intent Recognition (MIR) is a pivotal technology for understanding user psychology in human-computer interaction systems~\cite{paul2022intent, xu2019towards}, which aims to analyze linguistic, acoustic, and visual signals to accurately match user search content. Compared with single-modal recognition~\cite{chong2023leveraging, zou2022divide, zhang2021textoir}, MIR enables intelligent systems to capture more intricate intents, facilitating broad applications in multimedia retrieval~\cite{kofler2016user}, autonomous driving~\cite{kaffash2021big}, medical diagnosis~\cite{tiwari2022cnn, moon2022multimodal}, intelligent customer service~\cite{chen2019antprophet}, and voice assistants~\cite{dellaert2020consumer}. Previous research primarily relied on multimodal sentiment analysis (MSA) or dialogue act (DA) datasets, which lacked the granularity required for complex intent classification. To address this, MIntRec~\cite{zhang2022mintrec} and MIntRec2.0~\cite{zhang2024mintrec} were introduced as dedicated MIR datasets and established a more challenging benchmark for evaluating multimodal inconsistency.

Early MIR methods formulated initial baselines by adapting fusion methods from MSA, such as MulT~\cite{tsai2019multimodal}, MAG-BERT~\cite{rahman2020integrating}, and MMIM~\cite{han2021improving}. TCL-MAP~\cite{zhou2024token} was proposed as the first MIR method, with a series of task-specific approaches emerging subsequently. 
On the one hand, SDIF-DA~\cite{huang2023sdif} and InMu-Net~\cite{zhu2024inmunet} align multimodal features by enhancing cross-interaction, but their static fusion strategies hinder the model's ability to perform deep reasoning. Furthermore, while CAGC~\cite{sun2024contextual} and MuProCL~\cite{dong2025unbiased} use contrastive learning to improve semantic alignment, they overlook multi-view constraints on each signal, leading to over-reliance on the dominant modality. On the other hand, MISA~\cite{hazarika2020misa} and DuoDN~\cite{chen2024dual} attempt to decouple modality-invariant and specific representations, yet they remain at simple feature filtering without explicitly modeling semantic conflicts. Moreover, although MIntOOD~\cite{zhang2024multimodal} and MVCL-DAF~\cite{Hu_Zhang_Zhang_Ye_2025} dynamically adjust the modal contributions but lack the reliability evaluation, limiting model robustness. These methods have jointly driven the advancement of MIR, yet two core challenges remain: (1) the absence of a dual-pathway reasoning mechanism that enables adaptive cognitive regulation based on the sample conflict level; and (2) the failure to effectively model semantic inconsistency relations and conduct reliability evaluations.

To address these challenges, we propose the CDPR framework, which achieves a paradigm shift from low-level feature fusion to high-level cognitive reasoning. As illustrated in Figure~\ref{example}, textual and non-textual modalities often convey contradictory emotions (e.g., negative verbal content with positive non-verbal cues) in real-world scenarios. Existing methods frequently ignore this semantic inconsistency and fail to model complex semantic relations, leading to misclassification as \textit{Criticize}. In contrast, our method can adaptively regulate pathway weights based on conflict levels, effectively model inconsistency, and perform reliability evaluation, enabling it to correctly identify the intent as \textit{Joke}. This demonstrates that CDPR constructs a stable semantic foundation via the intuition pathway and mitigates high-level semantic conflicts through the reasoning pathway jointly, achieving efficient and robust intent understanding. Specifically, we design two parallel pathways to simulate the human thinking process. The intuition pathway rapidly captures multimodal consensus based on shared features to establish a solid semantic foundation. The reasoning pathway models semantic inconsistency based on private features and filters out conflicts with specific semantic patterns via learnable conflict vectors. It then corrects semantic features through statistical metrics to generate comprehensive conflict energy, mitigating high-level semantic conflicts and dynamically regulating the weights of the two pathways. Furthermore, to ensure the model learns structured multimodal features and prevents modality laziness, we introduce a multi-view loss function to supervise the learning process at various stages. Our contributions are summarized as follows:
\begin{itemize}[leftmargin=*]
    \item We propose a dual-pathway reasoning module that integrates consensus-based intuitive cognition with discrepancy-based deep reasoning to achieve intricate intent understanding.
    \item We introduce an inconsistency perception mechanism that quantifies fine-grained semantic conflicts via learnable vectors and assesses modality reliability with statistical metrics, achieving reliable multimodal conflict resolution.
    \item Extensive experiments conducted on two MIR datasets demonstrate that CDPR achieves SOTA performance, verifying its effectiveness and robustness.
\end{itemize}

\section{Related Works}
\subsection{Multimodal Intent Recognition}
Early research in multimodal understanding mainly relied on MSA datasets~\cite{zadeh2016mosi, zadeh2018mosei} and DA datasets~\cite{saha2020towards}. However, these lack the hierarchical intent labels in real-world scenarios. Subsequently, MIntRec~\cite{zhang2022mintrec} was introduced as the first dedicated MIR benchmark, providing 2,224 high-quality samples with 20 fine-grained intent categories, and initial baselines~\cite{tsai2019multimodal, rahman2020integrating, han2021improving} were constructed by transferring MSA approaches. Building on this, MIntRec2.0~\cite{zhang2024mintrec} expands the data scale and category diversity, establishing a challenging benchmark for intent understanding. To bridge the semantic gap between different modalities, contrastive learning has emerged as the dominant paradigm in the MIR field. TCL-MAP~\cite{zhou2024token} leverages video and audio to generate modality-aware prompts for token-level contrastive learning. CAGC~\cite{sun2024contextual} proposes a global contrastive objective for capturing dependencies across long video sequences. MVCL-DAF~\cite{Hu_Zhang_Zhang_Ye_2025} introduces multi-view contrastive learning to align heterogeneous features. Another line of research focuses on improving feature quality and refining interaction logic. InMu-Net~\cite{zhu2024inmunet} applies the information bottleneck principle to suppress noise and redundancy in non-verbal data. SDIF-DA~\cite{huang2023sdif} designs a shallow-to-deep interaction framework that progressively aligns features from coarse to fine granularity. WDMIR~\cite{ijcai2025p582} enhances nuanced semantic extraction by fusing video and audio features in the frequency domain. Furthermore, LGSRR~\cite{zhou-etal-2025-llm} utilizes MLLM to extract fine-grained semantics and construct logical reasoning relations. ARL~\cite{yangadaptive} alleviates modality imbalance by modeling modality importance at both the sample level and structural level.

\subsection{Multimodal Inconsistency Reasoning}
Multimodal data commonly exhibits inconsistency in real-world scenarios, which is primarily attributable to inter-modal semantic conflicts~\cite{hasan-etal-2019-ur, castro2019towards} or inherent data noise~\cite{huang2025mitigating, ma2023noisy}. Traditional approaches~\cite{zadeh2017tensor, liu2018efficient, yu2021learning} typically employed static fusion strategies that treat all modalities indiscriminately, leading to semantic cancellation rather than genuinely effective disambiguation. To mitigate this, recent research has branched into two main directions. One line of work leverages attention mechanisms to implicitly filter noisy modalities via dynamic weight allocation. For instance, TFR-Net~\cite{yuan2021transformer} utilizes self-attention to inversely map extracted high-level features back to the raw input space to guide the learning of complete semantics, while MIntOOD~\cite{zhang2024multimodal} employs a feature weighting network to adaptively assign importance to individual modalities. Another stream of research turns to feature disentanglement to alleviate inconsistency. For example, MISA~\cite{hazarika2020misa} projects each modality into modality-invariant and modality-specific subspaces to optimize representational consistency and distinctiveness. Similarly, DuoDN~\cite{chen2024dual} incorporates causal inference, employing a dual-pathway network with counterfactual intervention to disentangle semantic-oriented from modality-oriented representations. 3DGS~\cite{xiao2025multi} decouples inconsistencies across images from distinct perspectives by equipping each training view with a learnable grid, thereby significantly enhancing reconstruction quality.

\section{Methodology}
\subsection{Overview}
\begin{figure*}[t!]
	\centering  
	\includegraphics[scale=.67]{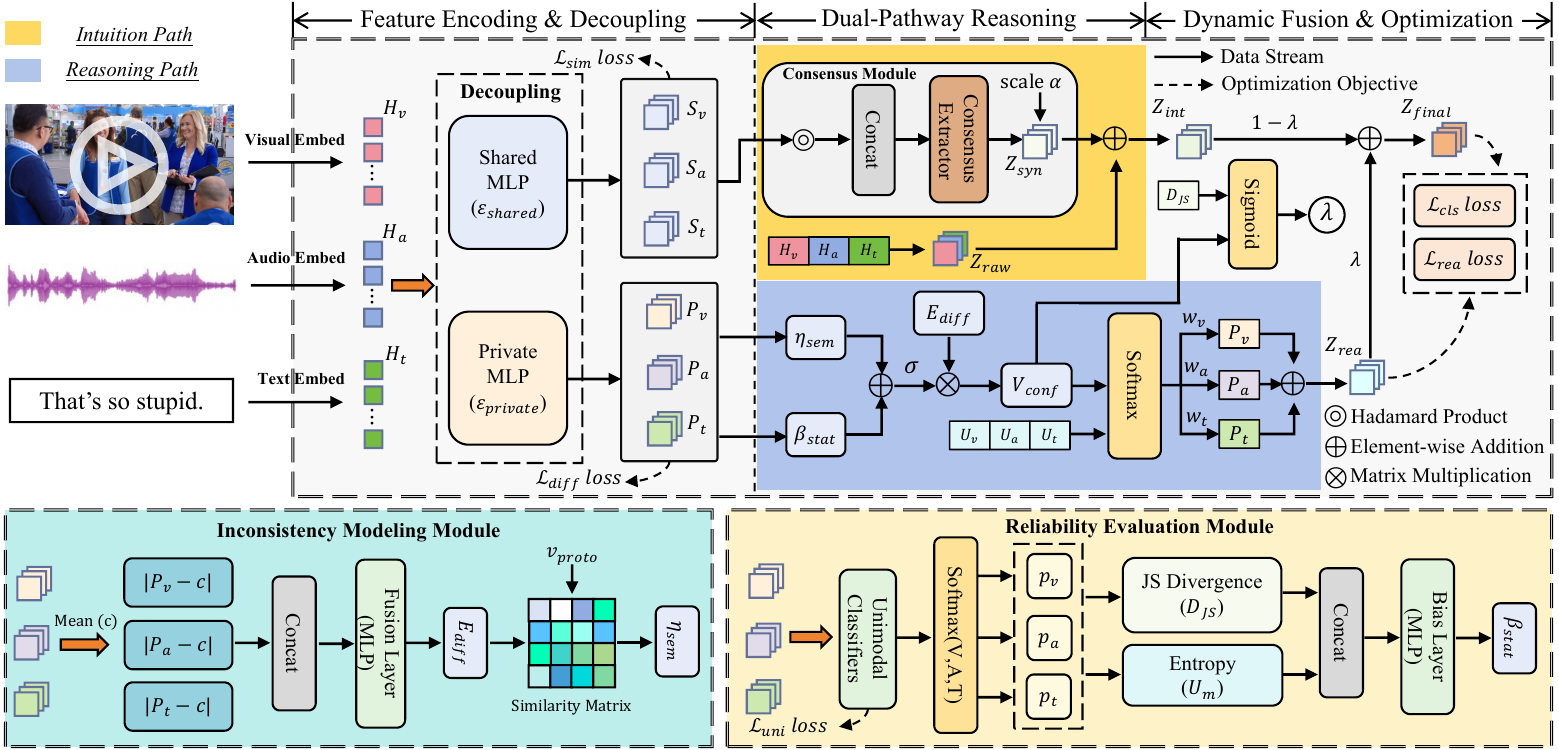}
	\caption{\label{model} The overall architecture of CDPR. Our approach comprises three key steps: (1) Dual-Pathway Reasoning, which integrates consensus-based intuitive cognition with discrepancy-based deep reasoning to model intricate semantic relations; (2) Inconsistency Perception Mechanism, which mitigates deep semantic conflicts through inconsistency modeling and reliability evaluation; and (3) Multi-View Loss Function, which supervises the training process at different stages.}
\end{figure*}

This section presents the framework of our Cognitive Dual-Pathway Reasoning (CDPR) method, as illustrated in Figure~\ref{model}. CDPR primarily consists of four components: Section~\ref{Feature Encoding and Decoupling} focuses on feature extraction and decoupling; Section~\ref{Cognitive Dual-Pathway Reasoning} dynamically regulates the intuition and reasoning pathways to model complex semantic relations; Section~\ref{Inconsistency Perception Mechanism} aims to perform inconsistency modeling and reliability evaluation; and Section~\ref{Multi-view Loss Functions} utilizes multi-view loss functions to supervise the learning process at different stages.

\subsection{Problem Definition}
\label{Problem Definition}
% This paper focuses on the Multimodal Intent Recognition (MIR) task. $\mathcal{D} = \{(X_i, y_i)\}_{i=1}^N$ denotes the dataset, where $y_i \in \mathbb{R}^C$ is the one-hot label vector for $C$ intent categories. For each sample, the input consists of three modalities: text, video, and audio, which can be represented as $X_t$, $X_v$, $X_a$. 
% Our objective is to learn a conflict-aware mapping function, which dynamically adjusts the intuition and reasoning pathways to jointly model deep semantic relations. Formally, the intuition pathway aims to capture fundamental modal consensus via shared features, while the reasoning pathway is dedicated to resolving high-level semantic conflicts. The final intent representation is modeled as an adaptively weighted sum of these two pathways, with the weights controlled by a global gating factor $\lambda$. This process is formalized as follows:
This paper focuses on the MIR task. $\mathcal{D} = \{(X_i, y_i)\}_{i=1}^N$ denotes the dataset, where $y_i \in \mathbb{R}^C$ is the one-hot label vector for $C$ intent categories. For each sample, the input consists of three modalities: text, video, and audio, which can be represented as $X_t$, $X_v$, $X_a$. 
Our objective is to learn a conflict-aware mapping function, which dynamically adjusts the intuition and reasoning pathways to jointly model deep semantic relations. Formally, the intuition pathway aims to capture fundamental modal consensus via shared features, while the reasoning pathway is dedicated to resolving high-level semantic conflicts. The final intent representation is modeled as an adaptively weighted sum of these two pathways, with the weights controlled by a global gating factor $\lambda$.
\begin{equation}
    \hat{y} = \mathcal{F}(X_t, X_v, X_a) \approx (1 - \lambda) \cdot \mathcal{F}_{int}(S) + \lambda \cdot \mathcal{F}_{rea}(P),
\end{equation}
where $\mathcal{F}_{int}$ and $\mathcal{F}_{reas}$ represent inference processes based on shared features and private features, respectively. $\lambda \in [0,1]$ is the global gating factor determined by the conflict level.

\subsection{Feature Encoding and Decoupling}
\label{Feature Encoding and Decoupling}
We first conduct shallow encoding on the text, video, and audio inputs, respectively. Aligning with prior research~\cite{zhang2022mintrec}, we adopt a standardized feature extraction pipeline. For the text modality, we utilize the powerful pre-trained language model BERT~\cite{devlin2019bert} to obtain text embeddings. For the video modality, frame-level visual features are extracted via the well-initialized Swin Transformer~\cite{liu2021swin}. For the audio modality, we employ the advanced speech recognition model WavLM~\cite{chen2022wavlm} to acquire signal-level audio features. We formulate this process as follows:
\begin{equation}
    H_t = \text{BERT}(X_t),
\end{equation}
\begin{equation}
    H_v = \text{SwinTransformer}(X_v),
\end{equation}
\begin{equation}
    H_a = \text{WavLM}(X_a),
\end{equation}
where $H_m \in \mathbb{R}^{l \times d}$ represents the feature vectors for each modality $m \in \{t,v,a\}$, with $l$ and $d$ denoting the sequence length and feature dimension, respectively.

To capture consistency signals and distinctive features between modalities, we follow the paradigm of MISA~\cite{hazarika2020misa}, adopting the feature decoupling strategy. Each modal feature $H_m$ is decomposed into a shared subspace $S_m$ and a private subspace $P_m$:
\begin{equation}
    S_m = \mathcal{E}_{shared}(H_m),
\end{equation}
\begin{equation}
    P_m = \mathcal{E}_{private}(H_m),
\end{equation}
where $\mathcal{E}_{shared}$ and $\mathcal{E}_{private}$ denote MLP encoding layers. The shared feature $S_m$ aims to capture modality-invariant features, while the private feature $P_m$ captures modality-specific features.

% model

\subsection{Cognitive Dual-Pathway Reasoning}
\label{Cognitive Dual-Pathway Reasoning}
Inspired by the Dual-Process Theory~\cite{kahneman2011thinking} in cognitive science, we design two parallel pathways to simulate the human thinking process. The intuition pathway rapidly captures multimodal consensus based on shared features $S_m$ to establish a stable semantic foundation. First, to preserve the complete contextual information of all modalities, we concatenate the raw features of the three modalities and project them through a non-linear mapping to obtain the multimodal contextual representation:
\begin{equation}
    Z_{raw} = \Phi_{raw}([H_t; H_v; H_a]),
\end{equation}
where $[\cdot; \cdot]$ denotes concatenation. Subsequently, to explicitly capture strong correlations between modalities, we calculate pairwise element-wise products for each modality to highlight consensus signals, yielding synergistic features:
\begin{equation}
    Z_{syn} = \Phi_{syn}([(S_t \odot S_v); (S_t \odot S_a); (S_v \odot S_a)]),
\end{equation}
% where $\odot$ denotes the element-wise product. Finally, we aggregate the contextual representation and synergistic features, introducing a learnable consensus scaling factor $\alpha$ to adaptively regulate the contribution of synergistic features, followed by layer normalization for stabilization:
% \begin{equation}
%     Z_{int} = \text{LayerNorm}(Z_{raw} + \alpha \cdot Z_{syn}).
% \end{equation}
% Furthermore, this residual structure allows the model to focus on raw information during the early stages and progressively shift focus towards more discriminative multimodal consensus patterns as training proceeds.
where $\odot$ denotes the element-wise product. Finally, we introduce a learnable scaling factor $\alpha$ to facilitate adaptive residual fusion. $\alpha$ is initialized to 0, allowing the model to primarily focus on contextual representation during the early stages of training, while progressively incorporating $Z_{syn}$ as an enhancement:
\begin{equation}
    Z_{int} = \text{LayerNorm}(Z_{raw} + \alpha \cdot Z_{syn}).
\end{equation}
This mechanism ensures training stability and enables the model to dynamically regulate the intensity of multimodal consensus signals based on sample characteristics.

% The text modality plays a dominant role in MIR tasks. However, most models tend to fall into text-centric shallow reasoning when facing intricate inconsistent intents, such as \textit{Joke}, \textit{Taunt}, or \textit{Criticize}, leading to erroneous intent predictions. Our method effectively mitigates semantic conflict by dynamically regulating the weights of each modality according to the Inconsistency Perception Mechanism (Section~\ref{Inconsistency Perception Mechanism}). Specifically, the reasoning pathway simulates the deep reasoning process when modality inconsistency arises. It operates on the private features $P_m$. The reasoning representation $Z_{rea}$ is a weighted sum of private features, where the weight $w_m$ represents the reliability score of modality $m$, generated dynamically based on semantic conflict and uncertainty:
The intuition pathway excels at establishing fundamental semantic consensus information, yet it fails to effectively address high-level inconsistent semantic relations. Therefore, we model semantic conflicts via the Inconsistency Perception Mechanism (Section~\ref{Inconsistency Perception Mechanism}), and the two pathways collaborate to achieve intricate intent understanding. Specifically, the reasoning pathway models semantic inconsistency and dynamically assigns a reliability score to each private feature, empowering the model to suppress modalities with high uncertainty while enhancing trustworthy cues.
\begin{equation}
    Z_{rea} = \sum_{m \in \{t,v,a\}} w_m \cdot P_m.
\end{equation}

The global gating factor $\lambda \in [0,1]$ reflects the balance between the two pathways, determined jointly by the semantic conflict vector and the discrepancy level:
\begin{equation}
    Z_{final} = (1 - \lambda) \cdot Z_{int} + \lambda \cdot Z_{rea}.
\end{equation}
A higher $\lambda$ indicates a higher level of conflict, prompting the model to assign more weight to the reasoning pathway.

\subsection{Inconsistency Perception Mechanism}
\label{Inconsistency Perception Mechanism}
The reasoning pathway aims to handle intricate semantic relations with multimodal inconsistencies. It avoids the blind fusion of all modal information, and instead dynamically evaluates the reliability of each modality through explicitly modeling semantic conflicts and statistical divergence, achieving deep mitigation.

In the private feature space, inconsistency between modalities manifests as geometric deviations of feature vectors. However, not all feature differences represent intent conflicts, which may stem from random noise or inherent modality heterogeneity. To capture genuine semantic conflicts, we first calculate the centroid $c$ of the three private features and the absolute deviation of each modality relative to this centroid:
\begin{equation}
    c = \frac{1}{3} \sum P_m, \quad \delta_m = |P_m - c|,
\end{equation}
where $|\cdot|$ denotes the element-wise absolute difference, and $\delta_m$ captures specific features in each modality that deviate from the multimodal consensus. To synthesize these deviations into a unified representation, we concatenate them and project to a hidden dimension $d$ via a non-linear transformation function $\Phi_{inc}$:
\begin{equation}
    E_{diff} = \Phi_{inc}([\delta_t; \delta_v; \delta_a]),
\end{equation}
where $[\cdot; \cdot]$ denotes concatenation along the feature dimension, yielding the difference vector $E_{diff}$. However, this vector aggregates all discrepancies, including noise, inherent differences, and genuine conflicts, which possess distinct geometric properties in high-dimensional space. Noise is isotropic, inherent differences are static biases, while genuine conflicts have structured directions related to labels. Therefore, we introduce a learnable conflict prototype vector $v_{proto}$ to distinguish conflicts with specific semantic patterns. The semantic conflict energy $\eta_{sem}$ is derived from the similarity between the difference vector and the prototype:
\begin{equation}
    \eta_{sem} = \text{Sim}(E_{diff}, v_{proto}) \cdot \tau,
\end{equation}
where $\tau$ represents the temperature parameter. A higher $\eta_{sem}$ implies that the current sample exhibits a semantic divergence pattern similar to the learned conflict prototype.

Although conflict energy reflects signal intensity, it fails to characterize signal reliability. To address this, we introduce a statistical bias as a decision-level calibrator. We first enable each modality to perform independent intent prediction, and then calculate the prediction probabilities for each class via the $\mathrm{Softmax}$ function to directly capture the inferential divergence across modalities:
% then measure the prediction probability for each class via $Softmax$ to directly perceive the prediction divergence among modalities:
\begin{equation}
    p_m = \text{Softmax}(\text{Classifier}_m(P_m)).
\end{equation}

Building on this, we quantify the overall divergence across modalities using Jensen-Shannon divergence~\cite{lin2002divergence}. The discrepancy level $D_{JS}$ is defined to measure the divergence between each unimodal distribution $p_m$ and its average distribution $p_{avg}$, where $D_{KL}$ denotes the Kullback-Leibler divergence~\cite{kullback1951information}:
\begin{equation}
    D_{JS} = \frac{1}{|M|} \sum_{m \in M} D_{KL}(p_m || p_{avg}).
\end{equation}

Furthermore, for highly uncertain intents predicted by a modality due to noise, their contribution should be reduced. Therefore, we introduce the Information Entropy~\cite{shannon1948mathematical} to quantify each modality’s uncertainty, which reflects the confidence of its independent prediction. High entropy corresponds to high confusion and low confidence, whereas low entropy indicates high confidence:
\begin{equation}
    U_m = -\frac{1}{\log C} \sum_{c=1}^{C} p_{m,c} \log p_{m,c},
\end{equation}
where $C$ denotes the total number of categories, and $p_{m,c}$ represents the probability that modality $m$ predicts the current sample belonging to the $c$-th category.

Subsequently, we concatenate the JS divergence $D_{JS}$ with the uncertainty vector $U_m$ to construct the statistical feature vector, which is projected through a learnable linear layer to obtain the statistical modulation bias $\beta_{stat}$:
\begin{equation}
    \beta_{stat} = W_{stat}[D_{JS}; U_t; U_v; U_a] + b_{stat}.
\end{equation}

Next, we aggregate the semantic conflict energy $\eta_{sem}$ with the statistical modulation bias $\beta_{stat}$ to calibrate the semantic judgment, yielding the comprehensive conflict energy $\eta_{conf}$:
\begin{equation}
    \eta_{conf} = \eta_{sem} + \beta_{stat}.
\end{equation}

Finally, we convert the comprehensive conflict energy into an attention gate between $[0,1]$ via the $\mathrm{Sigmoid}$ function and use it to perform weighted filtering on the original difference vector $E_{diff}$, obtaining the semantic conflict vector $V_{conf}$:
\begin{equation}
    V_{conf} = \sigma(\eta_{conf}) \cdot E_{diff}.
\end{equation}

The reliability scores $w_m$ for the reasoning pathway are generated by projecting the semantic conflict vector and uncertainty features, enabling the model to adaptively adjust weights based on conflict and confidence levels across modalities:
\begin{equation}
    w_{t,v,a} = \text{Softmax}([MLP_{trust}(V_{conf} + MLP_{unc}([U_t; U_v; U_a]))]).
\end{equation}

Finally, the global gating factor is jointly determined by the semantic conflict vector and the overall discrepancy level. It is mapped to $\lambda \in [0,1]$ via the $\mathrm{Sigmoid}$ function, determining the trade-off between the two pathways:
\begin{equation}
    \lambda = \text{Sigmoid}(\tanh(||V_{conf}||_2) + MLP_{con}(D_{JS})).
\end{equation}

\subsection{Multi-view Loss Functions}
\label{Multi-view Loss Functions}
To ensure the model learns structured multimodal representations, we adopt a multi-view joint optimization strategy to effectively supervise the model at various stages.

For the multi-grained supervision loss function, to prevent modality laziness caused by excessive reliance on a dominant modality and to ensure the reasoning pathway possesses independent discriminative ability during conflicts, we introduce auxiliary supervision at different granularities. Cross-entropy loss is applied to the final output $\mathcal{L}_{cls}$, the reasoning branch output $\mathcal{L}_{rea}$, and each unimodal auxiliary output $\mathcal{L}_{uni}$: 
\begin{equation}
    \mathcal{L}_{task} = \mathcal{L}_{cls} + \gamma_1 \mathcal{L}_{rea} + \gamma_2 \sum_{m} \mathcal{L}_{uni}(P_m, y).
\end{equation}

For the difference loss function~\cite{bousmalis2016domain}, to ensure private features capture modality-specific inconsistency cues, we enforce orthogonality between private and shared subspaces, as well as among different private subspaces, by minimizing this loss:
\begin{equation}
    \mathcal{L}_{diff} = \sum_{m} ||P_m^T S_m||_F^2 + \sum_{i \neq j} ||P_i^T P_j||_F^2.
\end{equation}

For the similarity loss function~\cite{zellinger2017central}, to extract cross-modal consensus, such as synchronized emotions in speech and facial expressions, we ensure the distribution alignment of shared features by minimizing the central moment discrepancy loss:
\begin{equation}
    \mathcal{L}_{sim} = \frac{1}{3} \sum_{i \neq j} \text{CMD}(S_i, S_j).
\end{equation}

Ultimately, the model is trained end-to-end by minimizing the weighted sum of the aforementioned losses.
\begin{equation}
    \mathcal{L}_{total} = \mathcal{L}_{task} + \beta_1 \mathcal{L}_{diff} + \beta_2 \mathcal{L}_{sim},
\end{equation}
where $\beta_1$ and $\beta_2$ are hyperparameters used to balance the contributions of task supervision and representation constraints.

\section{Experiments}
\subsection{Datasets}
To evaluate the effectiveness and robustness of CDPR for MIR, we conduct comprehensive experiments on two representative benchmark datasets:
(1) \textbf{MIntRec}~\cite{zhang2022mintrec} stands as the pioneering dataset for multimodal intent recognition, comprising 2,224 high-quality samples across 20 intent categories. It includes text, video, and audio modalities, with a split of 1,334 samples for training, 445 for validation, and 445 for testing;
(2) \textbf{MIntRec2.0}~\cite{zhang2024mintrec} is an extended version of MIntRec, which further expands the data scale and categorical diversity, consisting of 9,304 samples annotated with 30 fine-grained intent labels. 
It is divided into 6,165, 1,106, and 2,033 samples for training, validation, and testing, respectively.

% baselines
% \rowcolor{blue!15}
\begin{table*}[t!]
\centering
\caption{\label{main_results} Main results comparing CDPR with baselines on the MIntRec and MIntRec2.0 datasets.}
\resizebox{\linewidth}{!}{
\begin{tabular}{@{\extracolsep{2pt}}l|cccccc|cccccc}
    \toprule
     \multirow{2}{*}{Methods}& \multicolumn{6}{c|}{MIntRec} & \multicolumn{6}{c}{MIntRec2.0}\\
    &\makecell{ACC ($\uparrow$)} 
    & \makecell{F1 ($\uparrow$)} 
    & \makecell{P ($\uparrow$)} 
    & \makecell{R ($\uparrow$)} 
    & \makecell{WF1 ($\uparrow$)}
    & \makecell{WP ($\uparrow$)}
    &\makecell{ACC ($\uparrow$)} 
    & \makecell{F1 ($\uparrow$)} 
    & \makecell{P ($\uparrow$)} 
    & \makecell{R ($\uparrow$)} 
    & \makecell{WF1 ($\uparrow$)}
    & \makecell{WP ($\uparrow$)}\\
    \midrule
    MISA & 72.13 & 69.34 & 70.60 & 69.27 & 72.34 & 73.43 & 57.18 & 51.90 & 53.03 & 51.74 & 57.15 & 57.82 \\
    MULT & 71.69 & 67.71 & 68.78 & 67.76 & 71.38 & 71.79 & 58.58 & 51.64 & 54.74 & 51.49 & 57.43 & 57.49 \\
    MMIM & 71.73 & 68.92 & 69.51 & 69.21 & 71.60 & 72.21 & 56.35 & 50.48 & 52.52 & 50.85 & 55.44 & 56.44 \\
    MAG-BERT & 72.00 & 67.70 & 68.18 & 68.52 & 71.64 & 72.25 & 58.37 & 49.91 & 52.66 & 50.62 & 56.55 & 56.21 \\
    TCL-MAP & 73.35 & 69.31 & 69.50 & 70.30 & 72.92 & 73.27 & 57.83 & 52.16 & 54.02 & \underline{52.35} & 57.00 & 57.40 \\
    SDIF-DA  & 71.28 & 68.08 & 69.32 & 67.82 & 70.98 & 71.32 & 57.93 & 51.97 & 53.38 & 52.04 & 57.23 & 57.71 \\
    MIntOOD & 72.81 & 69.70 & 70.91 & 69.49 & 72.62 & 73.06 & 57.92 & 51.56 & \underline{56.91} & 51.29 & 56.75 & 58.67 \\
    MVCL-DAF & \underline{73.71} & \underline{70.33} & \underline{70.93} & \underline{70.44} & \underline{73.38} & \underline{73.57} & \underline{58.65} & \underline{52.27} & 56.50 & 51.53 & \underline{58.16} & \underline{58.96} \\
    \midrule
    CDPR & \textbf{75.15} & \textbf{71.04} & \textbf{72.01} & \textbf{71.08} & \textbf{74.91} & \textbf{75.37} & \textbf{60.82} & \textbf{53.86} & \textbf{57.88} & \textbf{53.40} & \textbf{59.54} & \textbf{60.23} \\
    $\Delta$ & 1.44$\uparrow$ & 0.71$\uparrow$ & 1.08$\uparrow$ & 0.78$\uparrow$ & 1.53$\uparrow$ & 1.80$\uparrow$ & 2.17$\uparrow$  & 1.59$\uparrow$ & 0.97$\uparrow$  & 1.05$\uparrow$  & 1.38$\uparrow$ & 1.27$\uparrow$\\
    \bottomrule 
\end{tabular}}
\end{table*} 

% ablation
\begin{table*}[t!]
\centering
\caption{\label{ablation_results} Ablation studies on MIntRec and MIntRec2.0 datasets. $\mathcal{P}$ and $\mathcal{L}$ denote pathway and loss function, respectively.}
\resizebox{\linewidth}{!}{
\begin{tabular}{@{\extracolsep{2pt}}l|cccccc|cccccc}
    \toprule
     \multirow{2}{*}{Ablation}& \multicolumn{6}{c|}{MIntRec} & \multicolumn{6}{c}{MIntRec2.0}\\
   
    &\makecell{ACC ($\uparrow$)} 
    & \makecell{F1 ($\uparrow$)} 
    & \makecell{P ($\uparrow$)} 
    & \makecell{R ($\uparrow$)} 
    & \makecell{WF1 ($\uparrow$)}
    & \makecell{WP ($\uparrow$)}
    &\makecell{ACC ($\uparrow$)} 
    & \makecell{F1 ($\uparrow$)} 
    & \makecell{P ($\uparrow$)} 
    & \makecell{R ($\uparrow$)} 
    & \makecell{WF1 ($\uparrow$)}
    & \makecell{WP ($\uparrow$)}\\
    \midrule
      w / o $\mathcal{P}_{int}$ & 73.39 & 69.71 & 70.19 & 70.40 & 73.56 & 74.58 & 59.04 & 52.42 & 56.64 & 52.50 & 58.27 & 59.35 \\
   w / o $\mathcal{P}_{rea}$ & 74.16 & 70.82 & 71.67 & 70.32 & 73.85 & 74.39 & 60.17 & 53.02 & 56.67 & 53.15 & 59.30 & \underline{60.19} \\
    w / o $\mathcal{L}_{sim}$ & \underline{74.21} & 70.56 & \underline{71.99} & 70.46 & \underline{74.20} & \underline{75.19} & 60.27 & 53.09 & 55.46 & 52.86 & 59.01 & 59.21 \\
    w / o $\mathcal{L}_{diff}$ & 74.20 & \underline{70.95} & 71.82 & 70.77 & 74.15 & 74.78 & 60.09 & 53.42 & \underline{57.69} & \underline{53.26} & 58.88 & 60.04 \\
   w / o $\mathcal{L}_{uni}$ & 73.33 & 70.72 & 71.45 & \underline{71.00} & 73.57 & 74.77 & \underline{60.57} & 53.36 & 56.14 & 53.23 & 59.32 & 59.67 \\
    w / o $\mathcal{L}_{rea}$ & 73.37 & 69.74 & 70.89 & 69.96 & 73.33 & 74.19 & 60.27 & \underline{53.71} & 56.67 & 52.91 & \underline{59.34} & 59.66 \\
    \midrule
    Full & \textbf{75.15} & \textbf{71.04} & \textbf{72.01} & \textbf{71.08} & \textbf{74.91} & \textbf{75.37} &  \textbf{60.82} & \textbf{53.86} & \textbf{57.88} & \textbf{53.40} & \textbf{59.54} & \textbf{60.23} \\
    \bottomrule 
\end{tabular}}
\end{table*} 

\subsection{Baselines and Evaluation Metrics}
We compared CDPR with eight baselines, from classic MSA models (MISA~\cite{hazarika2020misa}, MulT~\cite{tsai2019multimodal}, MMIM~\cite{han2021improving}, MAG-BERT~\cite{rahman2020integrating}) to advanced MIR methods (TCL-MAP~\cite{zhou2024token}, SDIF-DA~\cite{huang2023sdif}, MIntOOD~\cite{zhang2024multimodal}, and MVCL-DAF ~\cite{Hu_Zhang_Zhang_Ye_2025}). Detailed descriptions of these approaches are provided in Related Works and Introduction.

Following previous work~\cite{zhang2022mintrec}, we adopt the established evaluation metrics in MIR: accuracy (ACC), F1-score (F1), precision (P), recall (R), weighted F1-score (WF1), and weighted precision (WP) to evaluate our CDPR framework and baselines. For all metrics, higher scores indicate better performance.

\subsection{Implementation details}
In the experiments, the feature dimensions of text, video, and audio are 1024, 256, and 768, with the hidden layer dimension set to 768, while the sequence lengths of each modality are aligned using the CTC module~\cite{tsai2019multimodal}. The maximum training epoch is set to 40, with batch sizes 16, 8, 8 for training, validation, and testing. 
The early stopping strategy is configured with a patience of 5 epochs, and AdamW~\cite{loshchilov2018decoupled} is adopted as the optimizer, and the learning rate is tuned within the range [7e-6, 1e-5]. 
For MIntRec and MIntRec2.0 datasets, the warmup proportion is set to 0.05 and 0.01, and the temperature to 1.0 and 5.0, respectively, with dropout fixed at 0.2 and weight decay at 0.1. 
In addition, for the loss function coefficients, $\beta_1$ and $\beta_2$ are set to 0.1 for MIntRec and 0.01 for MIntRec2.0, while both $\gamma_1$ and $\gamma_2$ are fixed at 0.1. 
For fair comparison, all reported results are the average of five runs conducted on NVIDIA Tesla V100-SXM2 GPUs, using random seeds ranging from 0 to 4.

\subsection{Main Results}
Table~\ref{main_results} compares the performance between CDPR and competitive baseline methods. Our approach attains new SOTA results on both datasets across all metrics. On the MIntRec dataset, CDPR achieves an ACC of 75.15\% and WP of 75.37\%, surpassing the previous best performing baseline by 1.44\% and 1.80\%, respectively. Notably, while baselines like TCL-MAP and MVCL-DAF have attempted to improve fusion through contrastive learning or dynamic alignment, they still struggle with complex multimodal inconsistencies due to the lack of deep conflict modeling. In contrast, CDPR effectively mitigates this challenge via the cognitive dual-pathway mechanism, yielding comprehensive improvements in F1, P, R, and WF1 by 0.71\%, 1.08\%, 0.78\%, and 1.53\% respectively.

On the more challenging MIntRec2.0 dataset, the performance gain is even more significant. CDPR achieves an ACC of 60.82\% and an F1 of 53.86\%, outperforming the best competing method by 2.17\% and 1.59\%, respectively. Furthermore, CDPR maintains superiority across other metrics, yielding gains of 1.38\% and 1.27\% in WF1 and WP.
This notable improvement highlights the effectiveness of our approach in handling complex scenarios. In addition, we find that CDPR exhibits a more remarkable performance improvement of 8.43\% on MIntRec2.0, in contrast to the overall improvement of 7.34\% achieved on MIntRec, which indicates that our model possesses superior generalization capabilities and robustness when handling large-scale data with intricate modality conflicts.

% non-hard  hard
%  \small \footnotesize \scriptsize \tiny
\begin{table*}[t!]
\centering
\setlength{\tabcolsep}{10pt}
\caption{\label{hard_subsets} F1-score comparison between CDPR and baselines for intent categories on MIntRec dataset.}
\resizebox{16.0cm}{!}{
\begin{tabular}{@{\extracolsep{2pt}}l|cccc|cccc}
    \toprule
     \multirow{2}{*}{Methods}& \multicolumn{4}{c|}{Non-hard} & \multicolumn{4}{c}{Hard}\\
   
    & Praise & Apologise & Thank & Care & Joke & Taunt & Criticize & Oppose \\
    \midrule
    MISA & 86.63 & 97.78 & 98.03 & 87.14 & 38.74 & 22.15 & 53.44 & 36.15 \\
    MuIT & 84.72 & 97.93 & 96.83 & 88.12 & 33.95 & 26.12 & 49.72 & 34.68 \\
    MMIM & 86.78 & 97.66 & 96.76 & 87.38 & 37.23 & 26.58 & 50.15 & 31.20 \\
    MAG-BERT & 86.03 & 97.76 & 96.52 & 85.59 & 37.54 & 15.78 & 49.02 & 33.97 \\
    TCL-MAP & 87.20 & 97.70 & 97.00 & 86.80 & 29.00 & 17.20 & 51.30 & 35.90 \\
    SDIF-DA & 82.47 & 97.77 & 95.86 & 84.81 & 37.63 & 18.11 & 41.13 & 38.16 \\
    MIntOOD & 83.66 & 97.73 & 96.23 & 86.66 & 36.15 & 17.16 & 45.98 & 33.03 \\
    MVCL-DAF & 88.93 & 97.40 & 98.04 & 88.65 & 32.25 & 25.81 & 54.17 & 39.27 \\
    \midrule
    CDPR & \textbf{89.70} & \textbf{98.81} & \textbf{98.08} & \textbf{92.80} & \textbf{40.76} & \textbf{33.33} & \textbf{56.91} & \textbf{47.06} \\
    Human & 93.44 & 96.15 & 96.90 & 96.09 & 72.22 & 65.55 & 72.21 & 69.04 \\
    \bottomrule 
\end{tabular}}
\end{table*} 

% robustness analysis
\begin{table}[t!]
\centering
\caption{\label{robustness_analysis} Robustness analysis of CDPR and baselines on MIntRec with Gaussian noise injected into the text modality.}
\resizebox{\columnwidth}{!}{
\begin{tabular}{@{\extracolsep{2pt}}l|ccccc}
    \toprule
    \multirow{2}{*}{Methods} & \multicolumn{5}{c}{MIntRec} \\
    & $\sigma=0.0$ & $\sigma=0.1$ & $\sigma=0.3$ & $\sigma=0.5$ & $\sigma=0.7$ \\
    \midrule
    MISA & 69.34 & 63.70 & 49.26 & 28.11 & 11.75 \\
    MuIT & 67.71 & 62.47 & 48.02 & 28.41 & 12.37 \\
    MMIM & 68.92 & 63.78 & 53.57 & 33.42 & 17.98 \\
    MAG-BERT & 67.70 & 60.83 & 44.57 & 27.99 & 12.13 \\
    TCL-MAP & 69.31 & 64.09 & 51.37 & 31.05 & 14.12 \\
    SDIF-DA & 68.08 & 62.59 & 46.27 & 25.85 & 10.94 \\
    MIntOOD & 69.70 & 57.94 & 38.70 & 16.94 & 5.21 \\
    MVCL-DAF & 70.33 & 65.12 & 51.59 & 30.45 & 12.41 \\
    \midrule
    Ours & \textbf{71.04} & \textbf{67.84} & \textbf{55.65} & \textbf{36.34} & \textbf{22.68} \\
    \bottomrule 
\end{tabular}}
\end{table}

\subsection{Ablation Study}
To verify the contribution of each component in CDPR, we conducted comprehensive ablation studies on both MIntRec2.0 and MIntRec datasets, as summarized in Table~\ref{ablation_results}.

\textbf{Dual-Pathway Architecture:} (1) w/o $\mathcal{P}_{int}$: Removing the consistency module leads to the most significant performance drop with ACC falling by 1.78\% on MIntRec2.0 and 1.76\% on MIntRec. This indicates extracting modality-invariant consensus is fundamental for multimodal understanding. (2) w/o $\mathcal{P}_{rea}$: Removing the reasoning branch yields a clear decline with F1 dropping by 0.84\% on MIntRec2.0. This demonstrates that the reasoning pathway can effectively alleviate intricate semantic conflicts.

\textbf{Feature Decoupling:} (3) w/o $\mathcal{L}_{sim}$: Removing the similarity loss function causes ACC to drop by 0.94\% on MIntRec, validating that aligning shared feature distributions is a prerequisite for extracting reliable modality-invariant consensus. (4) w/o $\mathcal{L}_{diff}$: Removing the difference loss function results in a drop in F1 by 0.44\% on MIntRec2.0. This suggests that private features may become redundant with shared features without enforcing differences, weakening the ability of the reasoning pathway to capture unique modality-specific cues.

\textbf{Reasoning Supervision:} (5) w/o $\mathcal{L}_{uni}$: The removal of unimodal supervision leads to a noticeable drop with ACC declining to 73.33\% on MIntRec. This shows that this loss effectively prevents modality laziness, ensuring that each encoder learns discriminative features independently. (6) w/o $\mathcal{L}_{rea}$: The absence of direct supervision on the reasoning branch also impairs performance, confirming that deep supervision is necessary to force the reasoning pathway to learn effective conflict mitigation strategies.

\subsection{Performance on Hard Subsets}
To further investigate the effectiveness of the dual-pathway reasoning in handling diverse user intentions, we report the F1-scores for intent categories on MIntRec in Table~\ref{hard_subsets}. For Non-hard categories, where modalities are generally consistent, CDPR demonstrates SOTA performance, achieving 98.08\% on \textit{Thank} and 98.81\% on \textit{Apologise}, even surpassing human-level performance. This indicates that the intuition pathway successfully captures the multimodal consensus, building a solid semantic foundation without over-reasoning.
For hard categories that normally contain semantic contradictions, CDPR achieves a remarkable breakthrough, outperforming the strongest baseline by a substantial margin of 6.75\% on \textit{Taunt}. Similarly, CDPR delivers 47.06\% on \textit{Oppose}, surpassing MVCL-DAF by 7.79\%. These improvements are attributed to our reasoning pathway effectively mitigating semantic conflicts by modeling inconsistency and upweighting reliable modality signals. However, although our approach has made significant advances, the gap to human-level performance highlights the limitations in aligning multimodal features within existing models, inspiring us to develop more powerful and generalizable multimodal intent understanding systems.

% case study
\begin{figure*}[t!]
	\centering  
	\includegraphics[scale=.6]{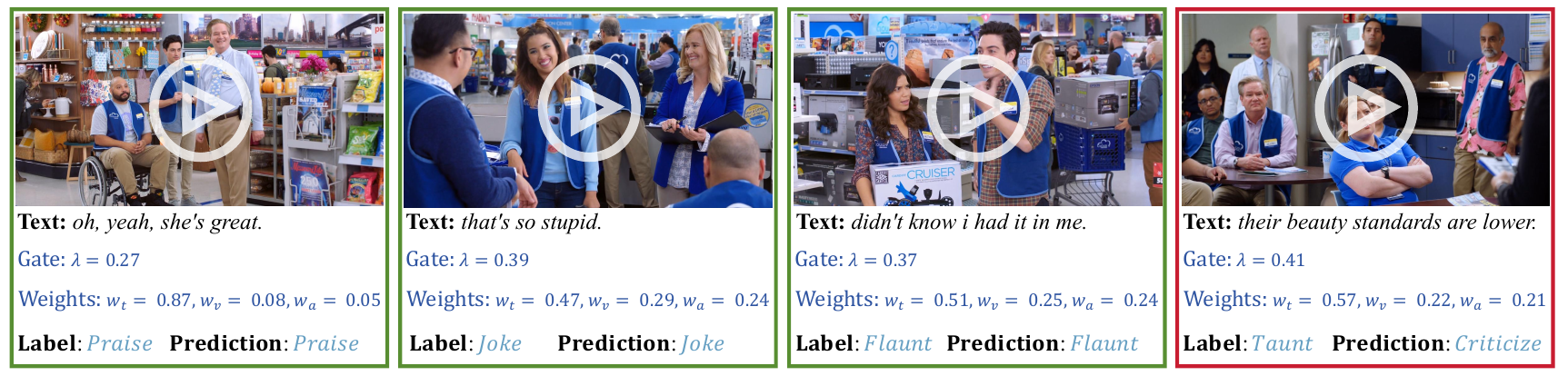}
	\caption{\label{case_Study} Qualitative analysis of representative samples from the MIntRec and MIntRec2.0 datasets.}
\end{figure*}

\subsection{Robustness Analysis}
To evaluate the reliability of CDPR in real-world scenarios, we conducted a robustness test by injecting Gaussian noise with varying intensities into the text input features on MIntRec, as shown in Table~\ref{robustness_analysis}, and our method achieves the most robust performance across all noise intensity levels. Specifically, at a moderate noise level with $\sigma=0.3$, CDPR maintains an F1 of 55.65\%, outperforming the second best method by 2.08\% and significantly surpassing other baselines. This indicates that our model can effectively filter out interference and retain critical semantic information. 
Most baselines suffer catastrophic performance drops under extreme noise conditions with $\sigma=0.7$, such as MIntOOD at 5.21\% and SDIF-DA at 10.94\%, while CDPR remains the most robust and achieves an F1 of 22.68\%. This confirms that existing methods heavily rely on text modality, whereas our approach adaptively protects the model from noise and ensures model stability in uncontrolled environments.
% This confirms that existing methods heavily rely on text modality, while our reliability evaluation module effectively filters out noisy inputs, adaptively protects the model from noise, and ensures model stability in unconstrained environments.

\subsection{Computational Complexity and Efficiency}
To evaluate the practical deployability of CDPR, we conducted a comprehensive efficiency analysis on the MIntRec2.0 dataset, as shown in Table~\ref{training_cost}. CDPR achieves SOTA performance without sacrificing computational efficiency. Compared to the strongest baseline MVCL-DAF, CDPR significantly reduces the parameter count by approximately 48\% and accelerates inference speed by around 123\%, from 33.69 samples/s to 75.18 samples/s. Remarkably, CDPR consumes merely 9 GB of GPU memory, making it the most hardware-friendly model among top performers, while MIntOOD requires nearly 24 GB. This highlights that our performance gains come from the ingenious design of the dual-pathway reasoning mechanism, rather than simple parameter stacking, making CDPR highly suitable for real-world deployment.

\subsection{Feature Distribution Visualization}
\begin{figure}[t!]
	\centering  
	\includegraphics[scale=.20]{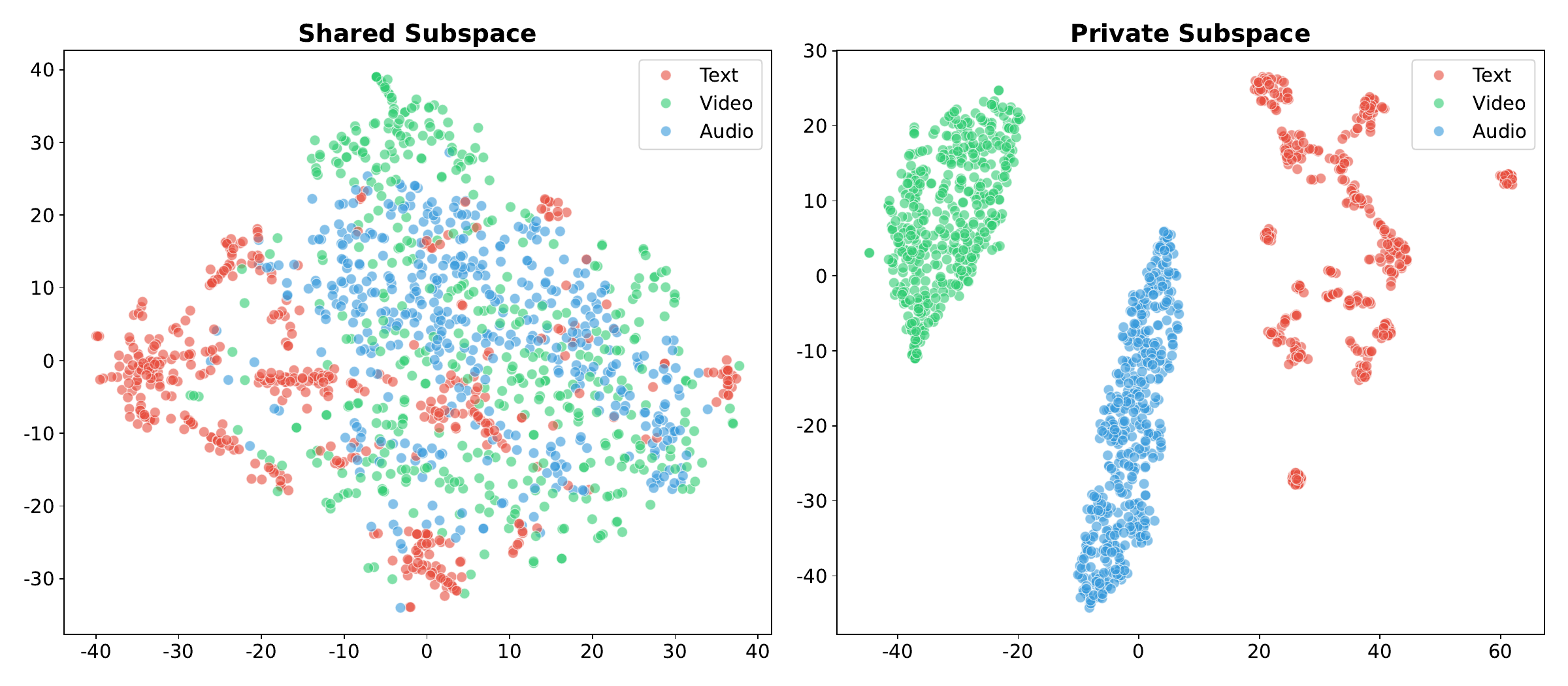}
	\caption{\label{t_sne} t-SNE visualization of feature distributions in the Shared and Private subspaces on the MIntRec dataset.}
\end{figure}

To intuitively verify the effectiveness of the representation decoupling strategy, we visualized the feature distributions of the shared and private subspaces using t-SNE on the MIntRec test set, as illustrated in Figure~\ref{t_sne}. The shared features from text, video, and audio modalities are uniformly mixed and indistinguishable, indicating that the similarity loss function has successfully aligned heterogeneous signals, compelling the encoder to extract modality-invariant consensus and enabling the intuition pathway to make rapid and consistent judgments. In contrast, private features form three well-separated clusters with clear boundaries between modalities. This indicates that the difference loss function effectively distinguishes unique characteristics. These distinct features are crucial for the reasoning pathway as they provide the distinctive signals required to detect conflicts and alleviate ambiguity.

\begin{table}[t!]
\centering
\caption{\label{training_cost} Comparison of computational efficiency between CDPR and baselines across four dimensions on MIntRec2.0.}
\resizebox{\columnwidth}{!}{
\begin{tabular}{@{\extracolsep{2pt}}l|cccc}
    \toprule
    \multirow{2}{*}{Methods} & \makecell{Params} & \makecell{GPU Mem} & \makecell{Training Time} & \makecell{IL} \\
    & \makecell{(M)} & \makecell{(MB)} & \makecell{(s/epoch)} & \makecell{(sample/s)} \\
    \midrule
    MIntOOD   & 344.14 & 24006 & 136 & 45.33  \\
    MVCL-DAF  & 669.51 & 17598 & 183 & 33.69  \\
    CDPR      & 346.80 & 9290 & 82 & 75.18  \\
    \bottomrule 
\end{tabular}}
\end{table}

% feature distribution visualization

\subsection{Case Study}
To clearly demonstrate the dual-pathway reasoning mechanism, we visualize the decision making process for four representative samples in Figure~\ref{case_Study}, including consistent, semantically conflicting, and error cases. For consistent samples, the video, audio, and text modalities all exhibit positive signals. Our model reduces the gating factor to $\lambda=0.27$ to prioritize the intuition pathway and actively relies on the text weights $w_t=0.87$, achieving efficient consensus information aggregation. In the inconsistent scenario, the model detects the conflict between negative text and positive video. It adaptively increases the reasoning pathway weight $\lambda$ to 0.39 and further suppresses the misleading text weight, dropping to 0.47, effectively correcting the prediction to \textit{Joke}. The error case indicates that strong text bias can sometimes overshadow subtle video cues when distinguishing fine-grained intents like \textit{Taunt} and \textit{Criticize}. These findings point to future research that captures nuanced non-textual features for discriminating highly correlated intents.

\section{Conclusion}
In this paper, we propose a novel CDPR framework that integrates consensus-based intuitive judgment with discrepancy-based deep reasoning to tackle the challenges of indiscriminate static fusion and the ineffective modeling of multimodal inconsistency. CDPR constructs a cognitive dual-pathway architecture that enables end-to-end unified processing ranging from the rapid processing of modality-invariant consensus to the deep resolution of modality-specific conflicts. 
Specifically, heterogeneous signals are first projected into shared and private subspaces via a representation disentanglement approach. Subsequently, the synergistic enhancement strategy of the intuition pathway is leveraged to swiftly aggregate consistent semantics, while the inconsistency perception mechanism effectively models semantic contradictions and performs reliability evaluation.
Crucially, it employs learnable prototype vectors and statistical metrics to model and quantify semantic conflicts, driving the model to dynamically adjust the weights of the intuition and reasoning pathways, achieving accurate reconstruction of complex semantic relations.
Extensive experiments conducted on MIR datasets demonstrate that the CDPR framework not only achieves SOTA performance but also exhibits stronger robustness and lower computational overhead, verifying its superiority in intricate multimodal interaction scenarios as well as its promising potential for practical deployment.

\begin{acks}
This work is supported by the Science Research Project of Hebei Education Department of China under Grant No. QN2024196. In addition, we would like to thank Professor Hua Xu and his team at Tsinghua University for their valuable support and assistance.
\end{acks}
% This work is supported by the Science Research Project of Hebei Education Department of China under Grant No. QN2024196 and Grant No. QN2026772. In addition, we would like to thank Professor Hua Xu and his team at Tsinghua University for their assistance.
% This work is supported by the Science Research Project of Hebei Education Department of China under Grant No. QN2024196 and Grant No. QN2026772. In addition, we would like to thank Professor Hua Xu and his team from the Department of Computer Science and Technology at Tsinghua University for their valuable support and assistance with this study.

%%
\bibliographystyle{ACM-Reference-Format}
\bibliography{references}

\end{document}